\documentstyle{article}

\begin{document}
\hspace{12cm}{\bf{ETH-TH/96-47}}

\hspace{12.5cm}{October 1996}

\vspace*{1cm}
\centerline{\large{\bf{Hamiltonian Reduction of
Diffeomorphism Invariant Field
  Theories}}}
\vspace{1cm}
\centerline{Jens Hoppe\footnote{Heisenberg Fellow. On leave of
    absence from Karlsruhe University.}}
\centerline{Theoretische Physik, ETH H\"onggerberg}
\centerline{8093 Z\"urich, Switzerland}
\vspace{0.5cm}
\centerline{and}
\vspace{0.5cm}
\centerline{Tudor Ratiu\footnote{Permanent
    address: Department of Mathematics,
    University of California, Santa Cruz,
    CA 95064, USA, Research partially
    supported by NSF Grant DMS-9503273
    and DOE contract DE-FG03-95ER25245-A000.}}
\centerline{Isaac Newton Institute for Mathematical Sciences}
\centerline{Cambridge University}
\centerline{20 Clarkson Road, Cambridge, CB3 0EH, U.K.}
\vspace{1cm}
\centerline{\large Abstract}
\vspace{0.5cm}
For a variety of diffeomorphism--invariant field theories describing
hypersurface motions (such as relativistic $M$-branes
in space-time dimension
$M+2$) we perform a Hamiltonian reduction
``at level 0'', showing that a
simple algebraic function of the normal velocity
is canonically conjugate to
the shape $\Sigma$ of the hypersurface.
The Hamiltonian dependence on $\Sigma$
is solely via the domain of integration,
raising hope for a consistent,
reparametrisation--invariant quantization.

\vfil\eject

As shown for relativistic membranes in 4 space--time
dimensions quite some
time ago [1], $M$--branes in $M+2$ dimensions
move such that in a frame and
parametrisation in which the hypersurface
moves normal to itself, the
energy--density is automatically
independent of time [2]. The
resulting equations of motion coincide
with those of a diffeomorphism
invariant Hamiltonian theory in which the
$M$ generators of spatial
diffeomorphisms are identically zero [3]
\footnote{
This is in agreement with a result proved in
[12] stating that in any diffeomorphism
invariant field theory that can
be coupled to gravity one must set the
momentum map to be zero (no other
value is allowed).}.
The Hamiltonian theory can be generalized to describe a large class
of hypersurface motions in Riemannian manifolds [4].
To describe
these in terms of diffeomorphism--invariant
variables while keeping their
Hamiltonian nature is highly desirable, especially
in view of the lack of a successful
reparametrization invariant quantization even
for strings (cp. [5], [6]).

The Hamiltonian reduction discussed in
this letter is astonishingly
simple. One finds that the unparametrized
hypersurface (i.e. its shape) is
conjugate to an algebraic function of
the normal velocity of the
hypersurface. This makes the reduced
Hamiltonian description simple, while
deepening the relation with fluid
dynamics found in [7] and [8]. Indeed,
in [9] and [10] the
Hamiltonian structure for dynamic free
boundary problems for 2 or 3 dimensional
incompressible homogeneous flows was determined,
substantially generalizing previous work
of Zakharov [11] on (incompressible homogeneous)
irrotational flows. Curiously, in the fluid
dynamical Poisson--bracket the shape of
the fluid's boundary (in [11], non-compact)
also appears as one of the
canonical variables. While the fluid motion
(reflecting the incompressibility) is volume-preserving
(our surface motions are not), a hidden relation between various
fluid--dynamical systems may nevertheless,
in view of [7], [8], also be
implied by the results of this letter.

\vspace{1cm}

The class of diffeomorphism--invariant
field theories for which we will
perform the Hamiltonian reduction ``at level 0''
is described by
\begin{equation}\label{Hamiltonian}
{\bf H} [\vec{x},\vec{p}] :=
\int_{\Sigma_0}d^M\varphi\sqrt{g}\,
h\left(\frac{p}{\sqrt{g}}\right)
\end{equation}
where $g$ is the determinant of the
the metric
$g_{rs}:=\frac{\partial\vec{x}}{\partial\varphi^r}
\cdot\frac{\partial\vec{x}}{\partial\varphi^s}$
induced by the usual inner product of ${\bf R}^{M+1}$
on the hypersurface $\Sigma$
defined by $\vec{x}:\Sigma_0\to\Sigma\subset{\bf R}^{M+1},~
p=\sqrt{\vec{p}^2}$ is the absolute value of the momentum vector,
and $h$ is a (monotonic) function of $w:=\frac{p}{\sqrt{g}}\ge 0$.
With $\vec{x}$ and $\vec{p}$ canonically conjugate,
the equations of motion derived from (\ref{Hamiltonian}) are:
\begin{equation}\label{canonical_equations}
\dot{\vec{x}} = h^{'}(w)\frac{\vec{p}}{p}\,, \qquad
\dot{\vec{p}} = \partial_r((h-h^{'}w)\sqrt{g} g^{rs}\partial_s\vec{x}).
\end{equation}
They imply the time independence of the functions
$C_r:=\vec{p}\cdot\partial_r\vec{x}, r=1 \dots m$, which generate
diffeomorphism of the reference-manifold
$\Sigma_0$. We will only be interested in
these constants of motion being identically zero, that is,
\begin{equation}\label{zero_level}
\vec{p}\cdot\partial_r\vec{x}\equiv 0 \quad\quad r=1,\dots, M \quad .
\end{equation}
In this case equations (\ref{canonical_equations}) become
\begin{equation}\label{motion_equations}
\dot{\vec{x}} = \pm h^{'}(w)\vec{n} \qquad
\dot{w} = \mp h(w)H \, ,
\end{equation}
while $w=\pm\frac{\vec{p}\cdot\vec{n}}{\sqrt{g}}$
($\vec{n}$ being the outward hypersurface normal, and
$H=-g^{rs}\partial^2_{rs}\vec{x}\cdot\vec{n}$
denoting the mean curvature of $\Sigma$), implying
\begin{equation}\label{first_equation}
\dot{v} = - h h^{''}H
\end{equation}
for the normal velocity $v:=\dot{\vec{x}}\vec{n} = \pm h^{'}(w)$.
Expressing $hh^{''}$ as a function of $v$, and viewing
$v$ as $\dot{\Sigma}$ (the rate of change of
$\Sigma$), (\ref{first_equation}) constitutes
the reparametrization--invariant
second order equation for the time--evolution of $\Sigma$.
Of course, the $\Sigma$--dependence of $H$ is non-trivial;
note e.g. that
\begin{equation}\label{H_equation}
\dot{H}=(-\triangle+R-H^2)v
\end{equation}
with $R$ and $\Delta$ being the Riemannian curvature
and the Laplacian on $\Sigma$. 
Denoting the normal component of $\frac{\vec{p}}{\sqrt{g}}$,
viewed as a function on $\Sigma$, by $u$ $ (\hat{=} \pm w)$,
as well as extending h to negative values of its argument by
defining $h(-w) = h(w)$ (which simultaneously removes the need to 
write $\pm $ in (4) and the unwanted constraint $w \ge 0$; $u$ is an 
unconstrained function on $\Sigma$, taking positive and negative 
values) (1) and (4) may be written as  
\begin{equation}\label{Hamiltonian1}
{\bf H} = \int_{\Sigma}h(u)\,,
\end{equation}
\begin{equation}\label{reduced_equations1}
\dot{\Sigma}=h^{'}(u), \quad \dot{u}=-hH.
\end{equation}
Noting that the first variation of the (hyper-)surface-area functional
equals the mean curvature, one is thus led to the fact that the Poisson bracket
on the reduced space is given by
\begin{equation}\label{Poisson_bracket}
\{F,G\}=\int_{\Sigma}\left (
\frac{\delta F}{\delta\Sigma}\frac{\delta G}
{\delta u}-\frac{\delta G}{\delta \Sigma}
\frac{\delta F}{\delta u} \right )\,,
\end{equation}
as the equations of motion derived via (9),
\begin{equation}\label{reduced_equations}
\dot{\Sigma} = \frac{\delta{\bf H}}{\delta u}\,, \qquad
\dot{u} = -\frac{\delta{\bf H}}{\delta\Sigma}\,\,,
\end{equation}
with $\bf H$ given by (7), (re)produce (8).

More generally, one can show that
\begin{equation}\label{var_der1}
\frac{\delta}{\delta\Sigma}\int_{\Sigma}f(\vec{x};u) =
Hf+\vec{n}\cdot\frac{\partial f}{\partial\vec{x}}
\end{equation}

\begin{equation}\label{var_der2}
\frac{\delta}{\delta\Sigma}\int_{\Sigma}n_kf(\vec{x};
u)=n_k\vec{n}\cdot\frac{\partial
f}{\partial \vec{x}}-
\frac{\epsilon_{ki_1\cdots i_M}}{(M-1)!}\,\,
n^{i_1}[f, x^ {i_2},\cdot , x^{i_M}]
\end{equation}
with $[f, x^{i_2},\cdots, x^{i_M}]$ defined on $\Sigma_0$ as
$\frac{\epsilon^{r_1\cdots r_M}}{\sqrt{g}} \,
\partial_{r{_1}}f\,\partial_{r_2}x^{i_2}\cdots\partial_r{_M}x^{i_M}$
($\epsilon^{r_1\cdots r_M}$ is the completely antisymmetric
symbol with $\epsilon^{1\cdots M}= 1$), making the last term in
(\ref{var_der2}) equal
to $\frac{d}{dx^k}f-n_k\vec{n}\cdot\frac{d}{d\vec{x}}f$, where
$\frac{d}{dx^k}=\frac{\partial}{\partial x^k}+\frac{\partial u}
{\partial x^k}\frac{\partial}{\partial u}; ~k=1,\cdots ,M+1$.
Formulae (\ref{var_der1}) and (\ref{var_der2})
follow from considering normal variations
$\vec{x}_{\epsilon}:=\vec{x}+
\epsilon\psi\vec{n}$ and calculating the coefficient of
$\sqrt{g}\psi$ in the integrand of
$\int_{\Sigma_0}\frac{d}{d\epsilon}|_{\epsilon=0}
(\sqrt{g_\epsilon}f_\epsilon\,\mbox{resp.}\,\sqrt{g_\epsilon}
f_\epsilon (\vec{n}_\epsilon)_k)$, keeping $u$ fixed; note that
$n_k=\frac{\epsilon_{ki_1\cdots i_M}}{M!}\,
[x^{i_1},\cdots, x^{i{_M}}]$
and $\frac{\epsilon_{ki_1\cdots i_M}}{(M-1)!} \,
[n^{i_1},x^{i_2}, \cdots , x^{i_M}]=Hn_k$.
Formulae (\ref{var_der1}) and (\ref{var_der2}) can be used to
verify that the generators of translation and rotation,
\begin{equation}\label{generators}
\vec{{\bf P}} = \int_{\Sigma}u\vec{n}
\quad {\rm and}\quad
{\bf L}_{ij} = \int_{\Sigma}u(x_in_j-x_jn_i)
\end{equation}
commute with ${\bf H}$. For the special case
\begin{equation}\label{Hamiltonian2}
{\bf H} = \int_{\Sigma}\sqrt{u^2+T^2}\quad ,
\end{equation}
describing relativistic $M$--branes in $M+2$ space
time dimensions ($T$ being
the dimensionful constant multiplying the $M+1$
dimensional world volume in the
original action) the symmetry is enhanced :
\begin{equation}\label{time_translation}
{\bf L}_{i0} := \int_{\Sigma} x_i\sqrt{u^2+T^2} \quad ,
\end{equation}
(13) and (14)
generate the inhomogeneous Lorentz group. One should
perhaps note that the second--order equations of motion
derived from
${\bf H}_{\alpha\beta}:=\int_{\Sigma}\sqrt{u^2+\alpha u+\beta}$,
\begin{equation}\label{special}
\dot{v}=-(1-v^2)H
\end{equation}
are independent of $\alpha$ and $\beta$ (actually, they
split into three different classes, according to
$(1-v^2)(u^2+\alpha u+\beta)=(\beta-\frac{\alpha^2}{4})$
being positive, zero, or negative). Finally note that
the hypersurface area ${\bf A}$ and the enclosed volume
${\bf V}$ evolve according to
\begin{eqnarray}\label{area_evolution}
\dot{{\bf V}} = \int_{\Sigma}h'(u) \quad {\rm and} \quad
\dot{{\bf A}} = \int_{\Sigma}Hh'(u) \, .
\end{eqnarray}

\vspace{1cm}

\centerline{\large Acknowledgement}
\vspace{0.5cm}
\noindent We would like to thank M. Bordemann and H.-H. Rugh
for useful discussions, J. Marsden for some very helpful
correspondence, and the Isaac Newton Institute for hospitality
and financial support.
\vspace{1cm}

\centerline{\large References}
\vspace{0.5cm}

\begin{description}
\item[[1]] J.~Hoppe; MIT Ph.D. Thesis 1982.
\vspace{-0.3cm}
\item[[2]] J.~Hoppe; hep-th/9503069.
\vspace{-0.3cm}
\item[[3]] J.~Hoppe; hep-th/9407103.
\vspace{-0.3cm}
\item[[4]] M.~Bordemann, J.~Hoppe; hep-th/9512001.
\vspace{-0.3cm}
\item[[5]] M.~L\"uscher, K.~Symanzik, P.Weisz; Nucl.~Phys. B 173 (1980) 365.
\vspace{-0.3cm}
\item[[6]] K.~Pohlmeyer; Phys.~Lett. B 119 (1982) 100.
\vspace{-0.3cm}
\item[[7]] M.~Bordemann, J.~Hoppe; Phys.~Lett.B 317 (1993) 315.
\vspace{-0.3cm}
\item[[8]] J.~Hoppe; Phys.~Lett. B 335 (1994) 41.
\vspace{-0.3cm}
\item[[9]] D.~Lewis, J.~Marsden, R.~Montgomery, T.~Ratiu; Physica 18D (1986)
  391.
\vspace{-0.3cm}
\item[[10]] A.~Mazer, T.~Ratiu; J.~Geom.~Phys. Vol.6 (1989) 271.
\vspace{-0.3cm}
\item[[11]] V.E.~Zakharov; J.~Appl.~Mech.~Techn.~Phys. Vol.9 \#2 1968.
\vspace{-0.3cm}
\item[[12]]  M.J.~Gotay and J.E. Marsden; Cont. Math. AMS. Vol 132 (1992)
  367.
\end{description}

\vspace{1cm}

\end{document}